\documentclass[graybox]{svmult}

\usepackage{mathptmx}       
\usepackage{helvet}         
\usepackage{courier}        
\usepackage{type1cm}        
\usepackage{makeidx}         
\usepackage{graphicx}        
\usepackage{multicol}        
\usepackage[bottom]{footmisc}

\makeindex             

\begin{document}

\title*{Structure and Cooling of Neutron and Hybrid Stars }
\author{S. Schramm, V. Dexheimer, R. Negreiros, T. Sch\"urhoff, J. Steinheimer}
\institute{S. Schramm \at FIAS, JW Goethe -Universit\"at, Ruth Moufang - Str. 1, D-60438 Frankfurt am Main, Germany \email{schramm@fias.uni-frankfurt.de}}
%
%
\maketitle

\abstract{
The study of neutron stars is a topic of central interest in the investigation of the properties of strongly compressed hadronic matter. 
Whereas in heavy-ion collisions the fireball, created in the collision zone, contains very hot matter, with varying density depending on  the beam energy, neutron stars largely sample the region of cold and dense matter with the exception of the very short time period of the existence of the proto-neutron star. Therefore, neutron star physics, in addition to its general importance in astrophysics, is a crucial complement to heavy-ion physics in the study of strongly interacting matter.
In the following, model approaches 
will be introduced to calculate properties of neutron stars  that incorporate baryons and quarks. These approaches are also able to describe the state of matter over a wide range of temperatures and densities, which is essential if one wants to connect and correlate star observables and results from heavy-ion collisions. The effect of exotic particles and quark cores on neutron star properties will be considered. In addition to the gross properties of the stars like their masses and radii their expected inner composition is quite sensitive to the models used. The effect of the composition can be studied through the analysis of the cooling curve of the star. In addition,
we consider the effect of rotation, as in this case the particle composition of the star can be modified quite drastically.}

\section{Introduction}

\label{sec:1}
A large number of experimental programs and theoretical efforts is devoted to the study of strong interaction physics under extreme conditions.
These conditions comprise large temperatures, densities, as well as extreme values of nuclear isospin.
In ultra-relativistic heavy-ion collisions a very hot fireball is created in the collision zone where hadronic matter is assumed to have melted into its
constituents, quarks and gluons. The net density in such reactions can be affected by the beam energy. At energies aimed for in the upcoming
FAIR facility at GSI a hot and also relatively dense system will be produced. 
On the other hand, in order to reach densities several times nuclear ground state density at relatively low temperatures, a study of the properties
of neutron stars is essential. In the following we will discuss a theoretical approach that is able to describe the conditions found in compact stars
as well as those created in heavy-ion collisions.

\section{Recent Observations}
\label{sec:2}

The main observational information about neutron stars is still the stellar mass that can help to constrain model descriptions of compact stars.
Here, a new benchmark has been set with the accurate measurement of the mass of pulsar PSR J1614-2230 of $M = 1.97 \pm 0.04\, M_\odot$ \cite{197} 
with several, much less certain, potential higher-mass candidates. A statistical analysis of measured masses suggests that there is no sign of a cut-off behavior 
of the mass distribution at the upper end, implying that higher values than 2 solar masses are plausible \cite{Kiziltan:2010ct}.

This new value serves to exclude a number of models or specific parameter sets, especially including hyperons, that have been in use before 
(see e.g. \cite{arXiv:1006.5660,arXiv:1107.2497, arXiv:1112.0234,Djapo:2008au}).

In the case of hybrid stars with a quark core in the center of the star, a quark phase based on a simple non-interacting quark model like the MIT bag model
also tends to reduce the maximum mass significantly (see the discussion in   \cite{arXiv:1011.2233}).


A quark phase that includes strong repulsive interactions, however, can have an equation of state quite similar to a nucleonic one, which avoids the softening of the matter and therefore the drop in maximum mass \cite{nucl-th/0411016,arXiv:1102.2869,arXiv:1108.0559}. Note, however, a potential problem with reproducing lattice QCD susceptibilities at small chemical potential in models with a strong repulsive quark-quark interaction \cite{arXiv:1005.1176}.

Another important measurement is the first observation of the real-time cooling behavior of a neutron star in the supernova remnant Cassiopeia A, 
where a rather steep drop of the surface temperature in the last 10 years has been recorded \cite{Heinke:2010cr}. This result has significant impact
on cooling studies of compact stars and might help to constrain the properties of matter in the interior of the star.

\section{The Quark-Hadron (QH) Model}
\label{subsec:2}
A long-lasting problem in modeling strong interaction physics originates from the fact that, depending on density $\rho_B$ and temperature $T$, the effective degrees of freedom
of QCD are completely different. Whereas at low values of $\rho_B$ and $T$ the world is hadronic, there is a transition to a deconfined and chirally symmetric system at some values of density and temperature. Whether this transition is a first- or higher-order phase transition or a smooth crossover is not known. The only relatively certain information on this point comes from lattice QCD simulations at zero chemical potential, indicating a cross-over transition. Although far from certain, at high densities and low temperatures one generally expects a first-order phase transition. If this is the case, somewhere in-between there should be at least one critical end-point of second order, the location of which is a prominent topic in heavy-ion research. 
A model description that should be valid over a large range of $T$ and $\rho_B$ (or, alternatively, the chemical potential $\mu_B$),
should be able to describe also a cross-over transition. Gluing together a hadronic and a quark equation of state (without extreme fine-tuning of parameters)
necessarily leads to a first-order transition over the whole range of $T$ and $\mu_B$. In order to avoid this and related problems
we developed a unified model of hadrons and quarks with the correct asymptotic degrees of freedom in the different regions of thermodynamic
parameters.

The hadronic part of the model is based on an effective chiral flavor-SU(3) model that includes the lowest SU(3) 
multiplets for baryons and mesons.  A detailed description of this general ansatz can
be found in \cite{Papazoglou:1997uw,Papazoglou:1998vr}. Restricting the discussion to the time-independent mean-field approximation the interaction of the baryons with the scalar and vector mesons 
reads  
\begin{equation}
L_{Int}=-\sum_i \bar{\psi_i}[\gamma_0(g_{i\omega}\omega+g_{i\phi}\phi+g_{i\rho}\tau_3\rho)+m_i^*]\psi_i ,
\label{coupling}
\end{equation}
summing over the baryon species $i$. $\omega$ and $\rho$ are the non-strange isovector 0 and 1 vector mesons, 
whereas $\phi$ denotes the vector meson consisting of an $s \bar{s}$ quark pair. The coupling to the scalar mesons is contained
in the effective baryon masses $m_i^*$:
\begin{equation}
m_i^* = g_{i\sigma} \sigma + g_{i\zeta} \zeta + g_{i\delta} \delta + \delta m_i ~~.
\end{equation}
The terms include the coupling of the baryons to the non-strange scalar isoscalar $\sigma$, isovector $\delta$ and strange fields $\zeta$.
In addition, there is a small explicit mass term $\delta m_i$.
Baryonic vacuum masses are generated by non-vanishing vacuum expectation values of the scalar mesons. These are the result of
the structure of the SU(3)-invariant scalar self-interactions, given by
\begin{eqnarray}
&L_{Self}= k_0(\sigma^2+\zeta^2+\delta^2)+k_1(\sigma^2+\zeta^2+\delta^2)^2+k_2\left(\frac{\sigma^4}{2}+\frac{\delta^4}{2}
+3\sigma^2\delta^2+\zeta^4\right)\nonumber\\&
+k_3(\sigma^2-\delta^2)\zeta+k_4\ \ \ln{\frac{(\sigma^2-\delta^2)\zeta}{\sigma_0^2\zeta_0}}~.&
\end{eqnarray}
In addition, the full Lagrangian includes self-interactions of the vector mesons and
a further term that breaks chiral symmetry explicitly generating the masses of the pseudo-scalar mesons. A detailed discussion of the model and specific values
of the parameters can be found in \cite{Dexheimer:2008ax}.
 
\begin{figure}[th]
\centerline{\includegraphics[width=6.4cm]{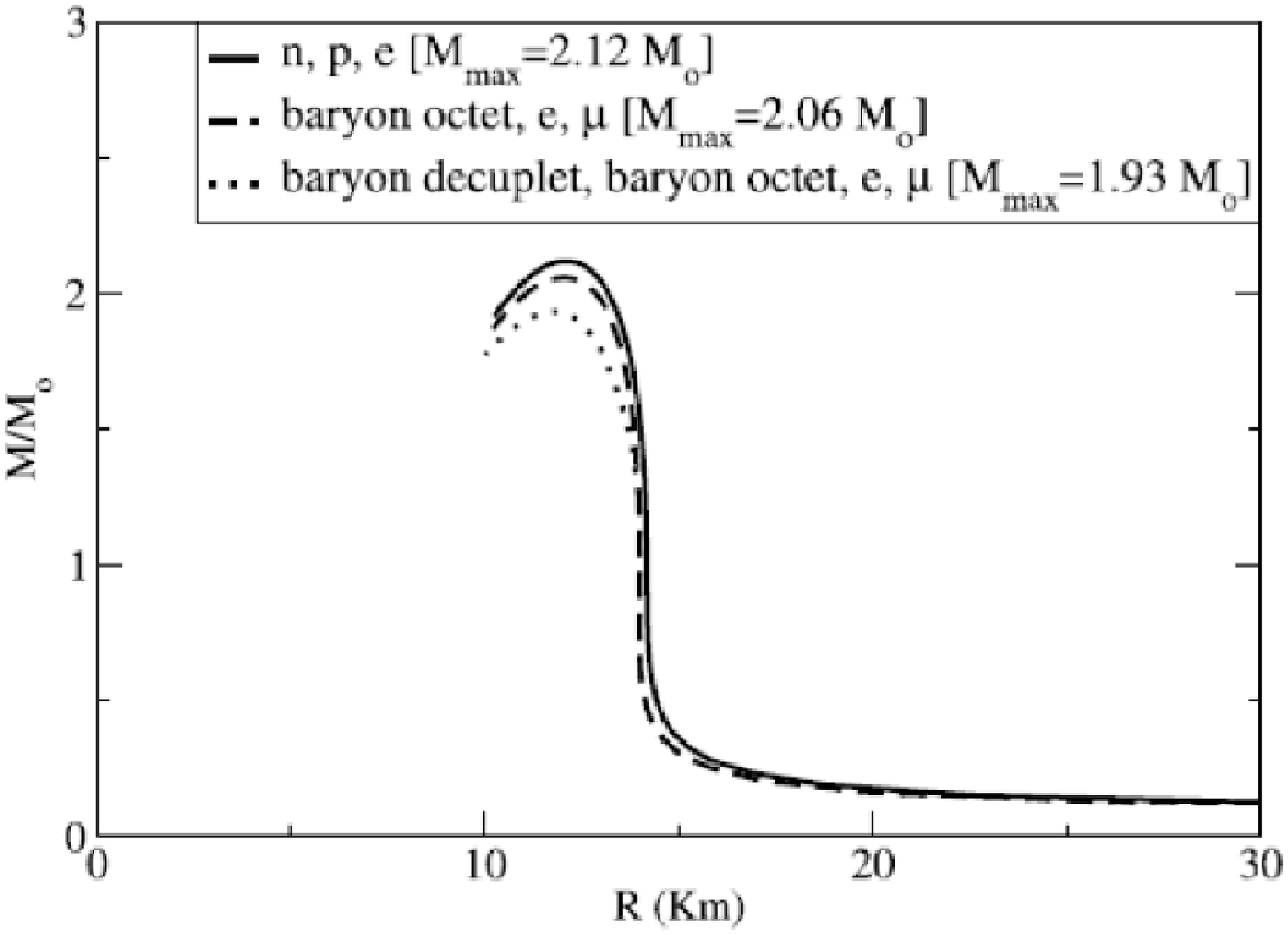}\includegraphics[width=5.8cm]{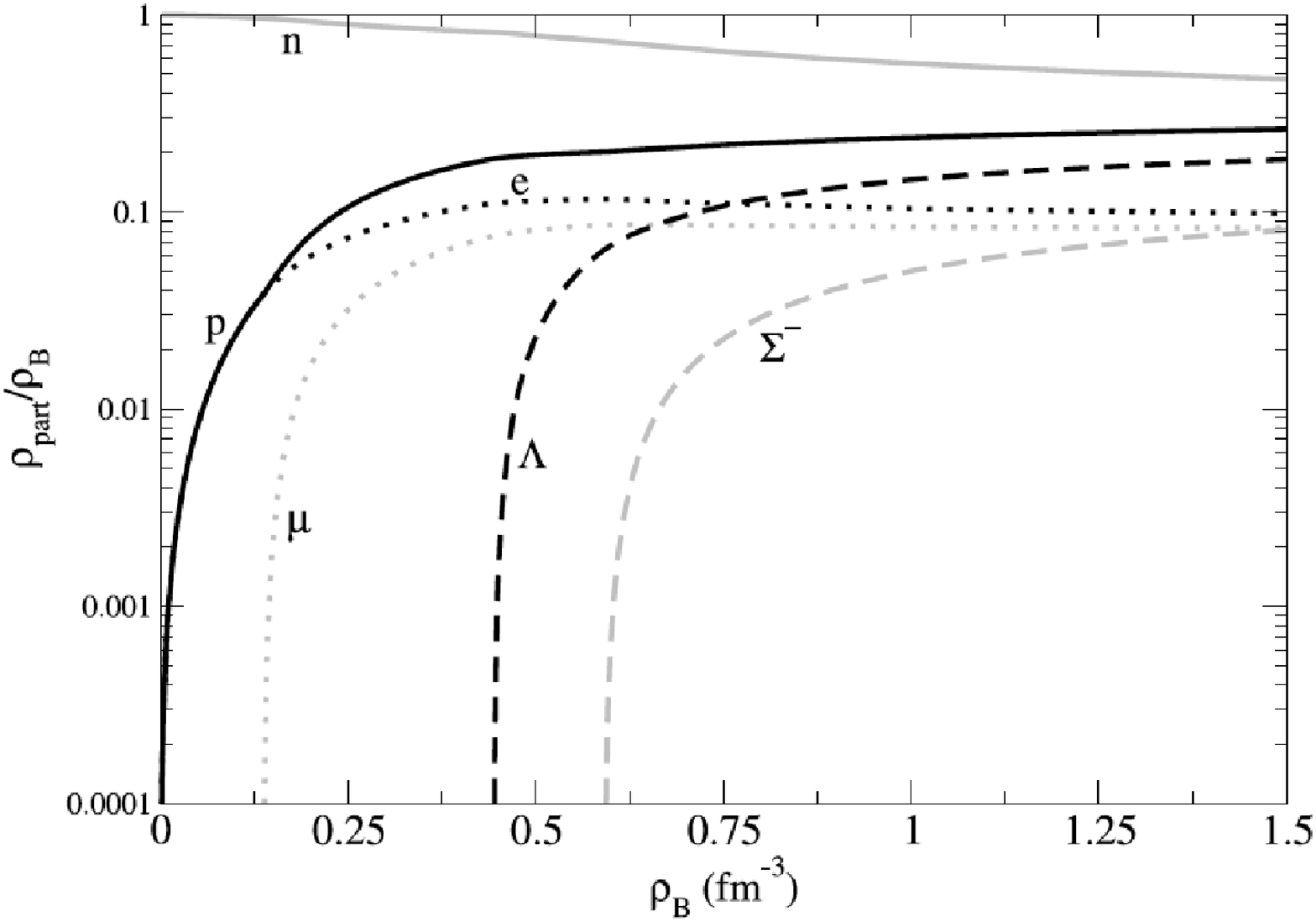}}
\vspace*{5pt}
\caption{Left: Star masses as function of radii including nucleons, hyperons, and the baryonic spin 3/2 decuplet \protect\cite{Dexheimer:2008ax}.
Right: Relative particle populations as function of baryon density including the lowest baryon octet.}
\label{mr}
\end{figure}

\begin{figure}[th]
\centerline{\includegraphics[width=6.4cm]{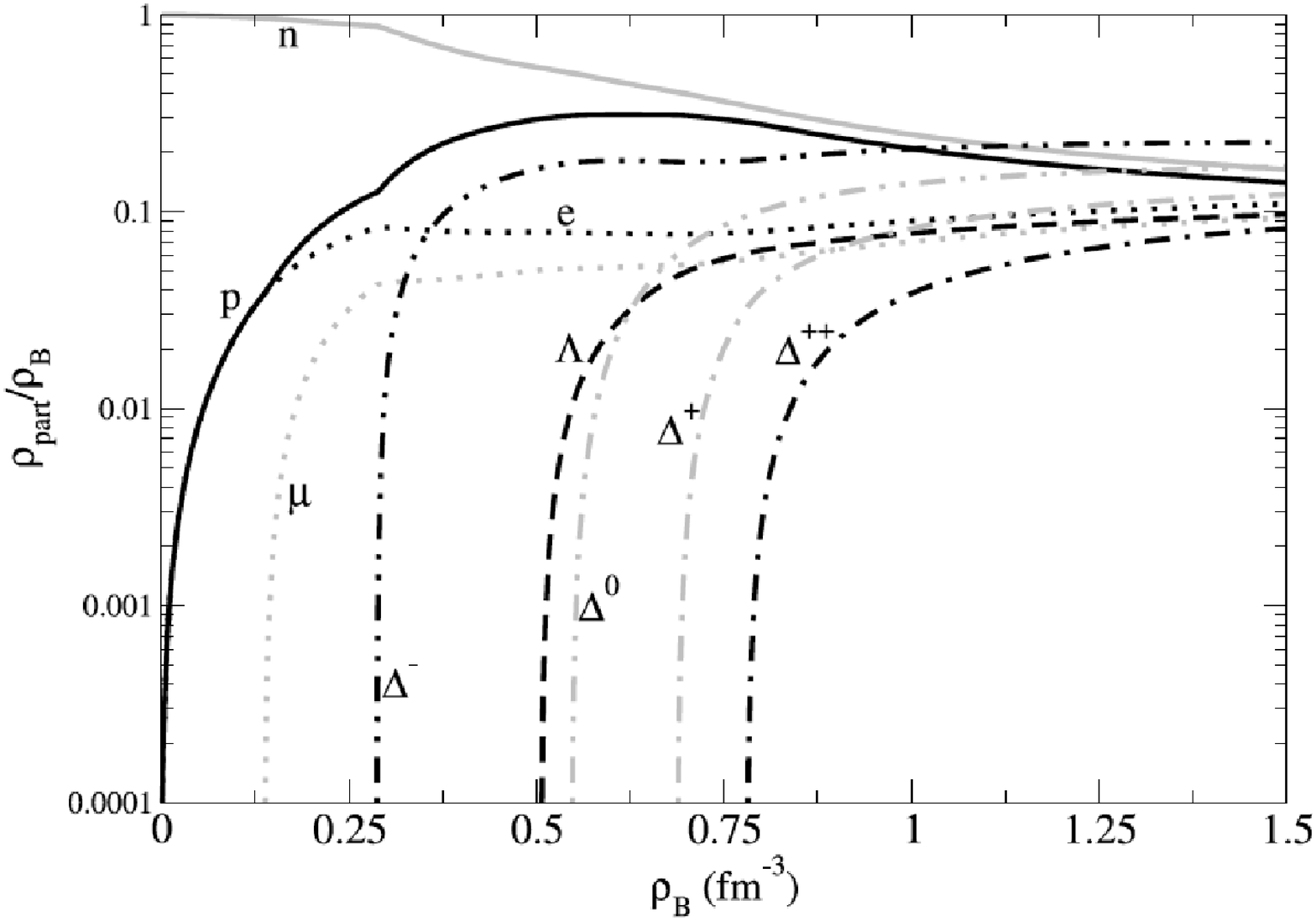}\includegraphics[width=5.8cm]{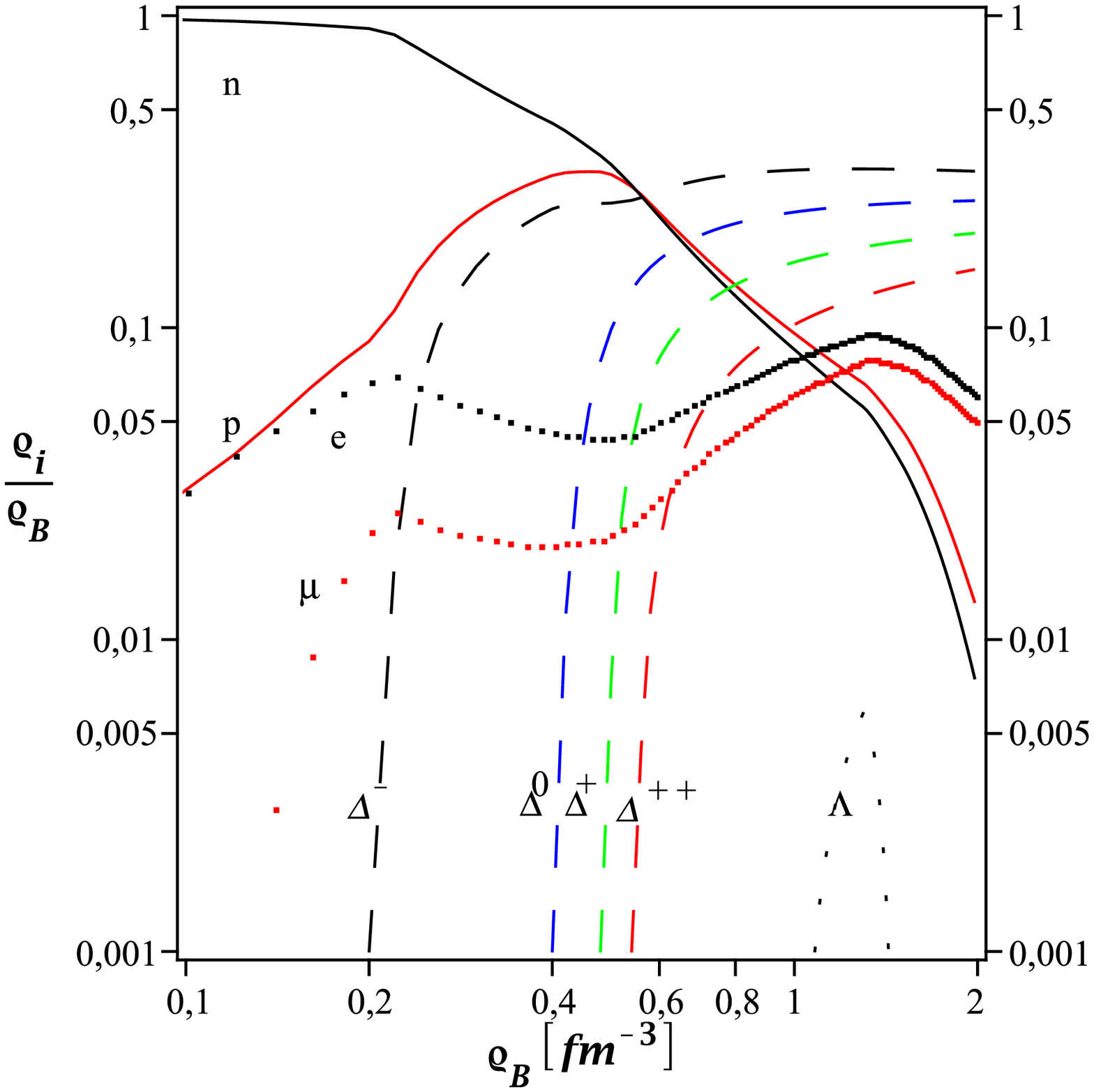}}
\vspace*{5pt}
\caption{Left: Relative particle densities as function of baryon density including the spin-3/2 decuplet. 
Right: Same as the left panel for a relative
strength of the $\Delta-\omega$ coupling $r_V = g_{N\omega}/g_{\Delta\omega} = 0.9$. 
}
\label{popstar}
\end{figure}

\begin{figure}[th]
\centerline{\includegraphics[width=6.4cm,height=5cm]{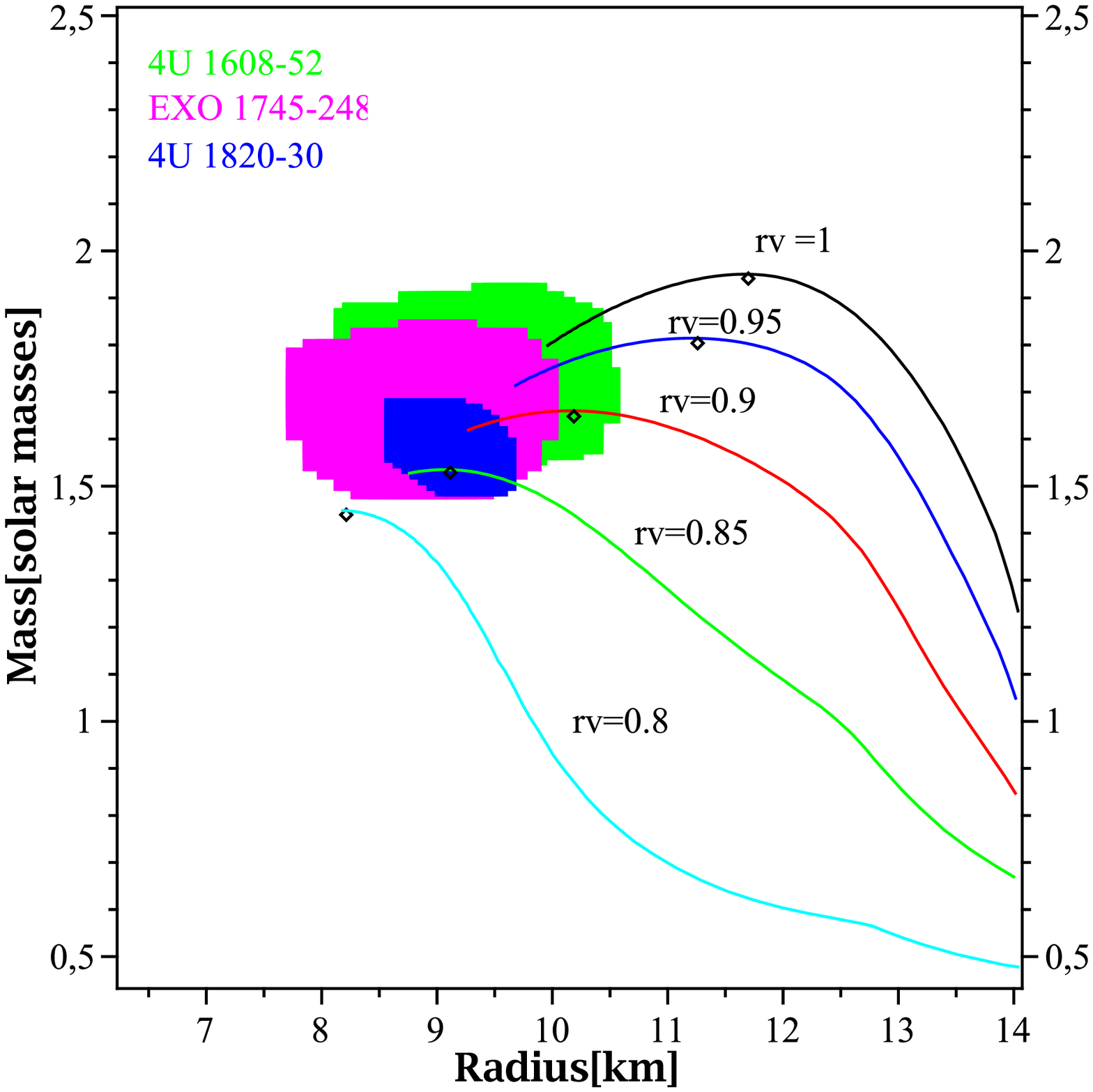}\includegraphics[width=5.8cm]{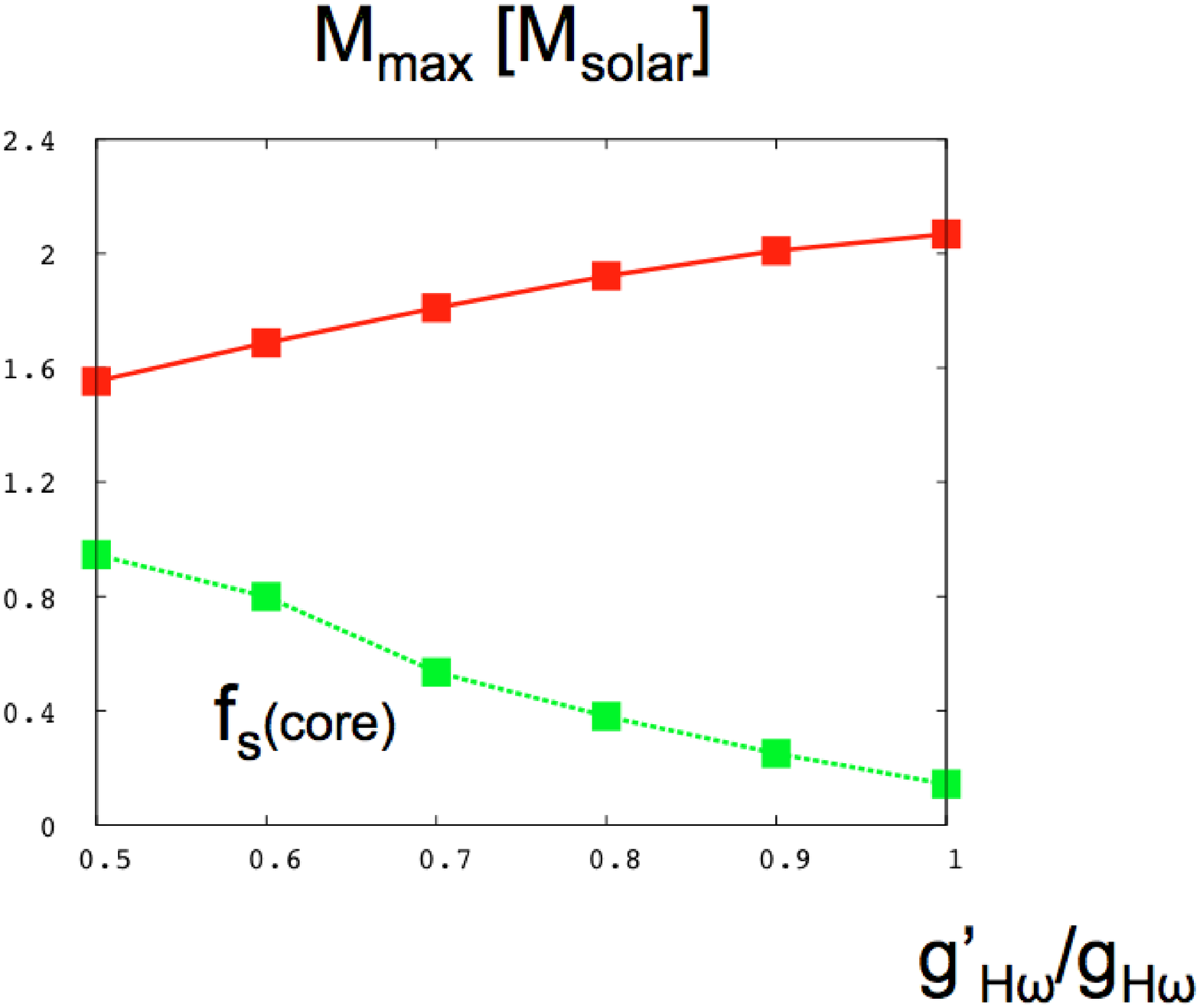}}
\vspace*{5pt}
\caption{Left: Mass-radius diagrams for different values of $r_V$. Also shown are the results of an analysis of the probable value range of mass and radius for the cases of three
low-mass X-ray binaries \cite{Ozel:2010fw}.
Right: Maximum star mass and strangeness content $f_s$ of the star as function of the strength of the vector repulsion of the hyperons.}
\label{dmr}
\end{figure}

 In order to study compact stars we determine the equation of state within this model and solve the
 Tolman-Oppenheimer-Volkoff equations for static stars \cite{tov1,tov2}. The resulting  star masses and radii of stars are shown in Fig. \ref{mr}.
As could be expected,
including hyperons in addition to nucleons, and further taking into account the baryonic spin 3/2 decuplet (mainly, the $\Delta$ resonances) reduces the maximum mass of the compact stars from 2.12 to 1.93 solar masses
as more degrees of freedom translate to a softer equation of state. However, the influence of the hyperons is rather weak with a value of $f_s$, the amount of 
strangeness per baryon, of about $0.1$ in the core of the heaviest star. 
Note, that the internal structure of the stars can change quite considerably, although the maximum masses drop by only 10 percent.
This can be seen by comparing Fig. \ref{mr} (right panel) and  Fig. \ref{popstar} (left panel). Including the $\Delta$ resonances changes the population 
of particles substantially, 
removing the $\Sigma^-$ and suppressing the $\Lambda$ hyperon,
replacing them by the corresponding $\Delta$ states $\Delta^-$ and $\Delta^0$. 
In these calculations it was assumed that the baryon decuplet has the same vector meson coupling
strengths as the octet, that is, for instance in the case of the $\Delta$, $g_{N\omega} = g_{\Delta\omega}$. Changing this value moderately, by reducing the $\Delta$ coupling, can alter
the results significantly as it was investigated in \cite{Schurhoff:2010ph}. 
The results are shown in Fig. \ref{popstar}. In the right panel of the figure the particle densities are shown for a ten percent reduction of $g_{\Delta\omega}$.
It is interesting to note that for densities beyond four times saturation density the $\Delta$'s dominate the system compared to nucleons.
This parameter adjustment, however, has a sizable effect on  masses and radii of the stars as can be seen in Fig. \ref{dmr} (left panel). Also shown are possible ranges for masses
and radii of three X-ray binaries as analyzed  in \cite{Ozel:2010fw}. By lowering the coupling one can reproduce the results of the experimental analysis, but the maximum masses are
below the observed value of two solar masses.
 
In a similar way one can investigated the influence of the hyperons on the results.
The rather small influence of the hyperons on the star properties is largely due to the meson interactions that keep the strange scalar field relatively large at high densities. Thus hyperons stay heavy and are not very strongly populated  \cite{Dexheimer:2008ax}.
Fig. \ref{dmr} (right panel) shows the result of a calculation using the same model but reducing by hand the vector coupling constant of the hyperons at densities beyond nuclear matter densities (thus without changing the reasonably well-known optical potential depths in normal nuclear matter that are reproduced in the model). One can see that a reduction of the coupling strength by 50 percent increases the strangeness fraction  in the core of the star ten-fold from 0.1 to 1, which corresponds
to an average of one strange quark per baryon. This softens the equation of state significantly, and the maximum mass is reduced by half a solar mass.

\begin{figure}[th]
\centerline{\includegraphics[width=5.4cm,height=4.5cm]{figs/Mass2.eps}\includegraphics[width=6.4cm]{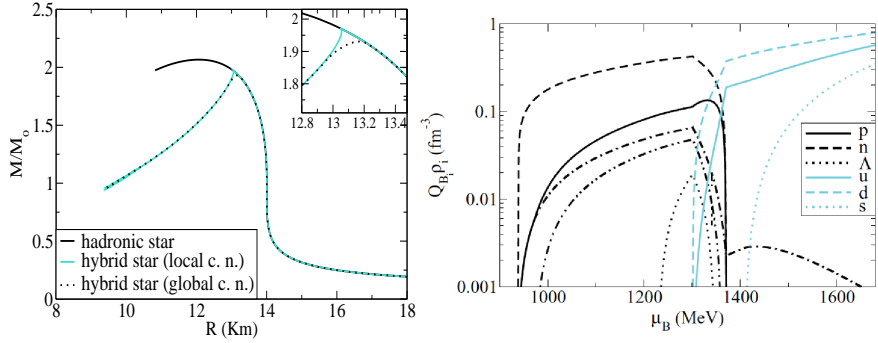}}
\vspace*{5pt}
\caption{Left: 
Mass-radius diagram of the quark-hadron model compared to the purely baryonic case. The inset shows the effect of introducing a Gibbs mixed phase.
Right: Particle population in the star as function of baryon density. Note that there are practically no strange particles in the interior of the star.}
\label{Mgh}
\end{figure}

Following the ideas of the PNJL approach \cite{Fukushima:2003fw,Ratti:2005jh} we include quark degrees of freedom  and an effective field $\Phi$, in 
analogy to the Polyakov loop, describing the 
deconfinement phase transition.
Here, the quark fields couple linearly to the scalar and vector condensates as in Eq. (\ref{coupling}).
$\Phi$ couples to the hadron and quark
masses such that quarks attain a high mass in the confined phase at low values of $\Phi$ and correspondingly hadrons obtain a large mass for large values of the field, removing the baryons as degrees of freedom.
The values of the parameters are quoted in \cite{Dexheimer:2009hi,Negreiros:2010hk}.
In addition to the usual structure of the effective potential of the field $\Phi$ we add chemical-potential dependent terms.
They can be chosen to reproduce the position of the critical end point of a first-order phase transition line 
as suggested by lattice calculations \cite{Fodor:2004nz}.
The  star masses and radii get modified due the quark contributions. The results are shown in the left panel of Fig. \ref{Mgh}.
The quarks largely cut off the branch of stable masses with a maximum value of  $M = 1.93\,M_\odot$.
A Gibbs phase mixture in the core of the star leads to a 2 km mixed region of quarks and baryons.

In a different theoretical approach to chirally symmetric models we studied the chiral symmetry restoration in the so-called parity-doublet 
model \cite{Detar:1988kn,Dexheimer:2007tn,Dexheimer:2008cv}.
Here one extends the baryonic states by the (hypothetical) partner states with opposite parity. In the case of the nucleon one candidate is
the N(1535) resonance. In this formulation the signal for chiral symmetry restoration is given by the degeneracy of the parity partners,
in analogy to the scalar and pseudo-scalar mesons in the linear $\sigma$ model, which obtain equal masses at high temperature..
Extending this approach to the whole SU(3) octet \cite{Steinheimer:2011ea} the equations look similar to the ones above with the exception of the effective baryon
masses that read:
\begin{equation}
m^*_{i\pm} = \sqrt{ \left[ (g^{(1)}_{\sigma i} \sigma + g^{(1)}_{\zeta i}  \zeta )^2 + (m_0+n_s m_s)^2 \right]}
\pm g^{(2)}_{\sigma i} \sigma \pm g^{(1)}_{\zeta i} \zeta ,
\end{equation}
where $\pm$ denote the parity partners, $m_s$ is the strange quark mass and $n_s$ is given by the number of strange quarks in the corresponding baryon.
Due to the doublet structure there are now twice as many scalar coupling constants. From this expression one can see that in the case of vanishing
fields $\sigma$ and $\zeta$, the particle masses of the parity partners are degenerate with a mass of the parameter $m_0$, in the case of nucleons.
A full discussion of the model is given in \cite{Steinheimer:2011ea}.
The 	quarks are again coupled to the meson fields. Here we take into account an excluded volume correction for the hadrons in a thermodynamically
consistent formulation, which automatically leads to a switch to quarks at high temperature and/or densities.

\begin{figure}[th]
\centerline{\includegraphics[width=5.8cm]{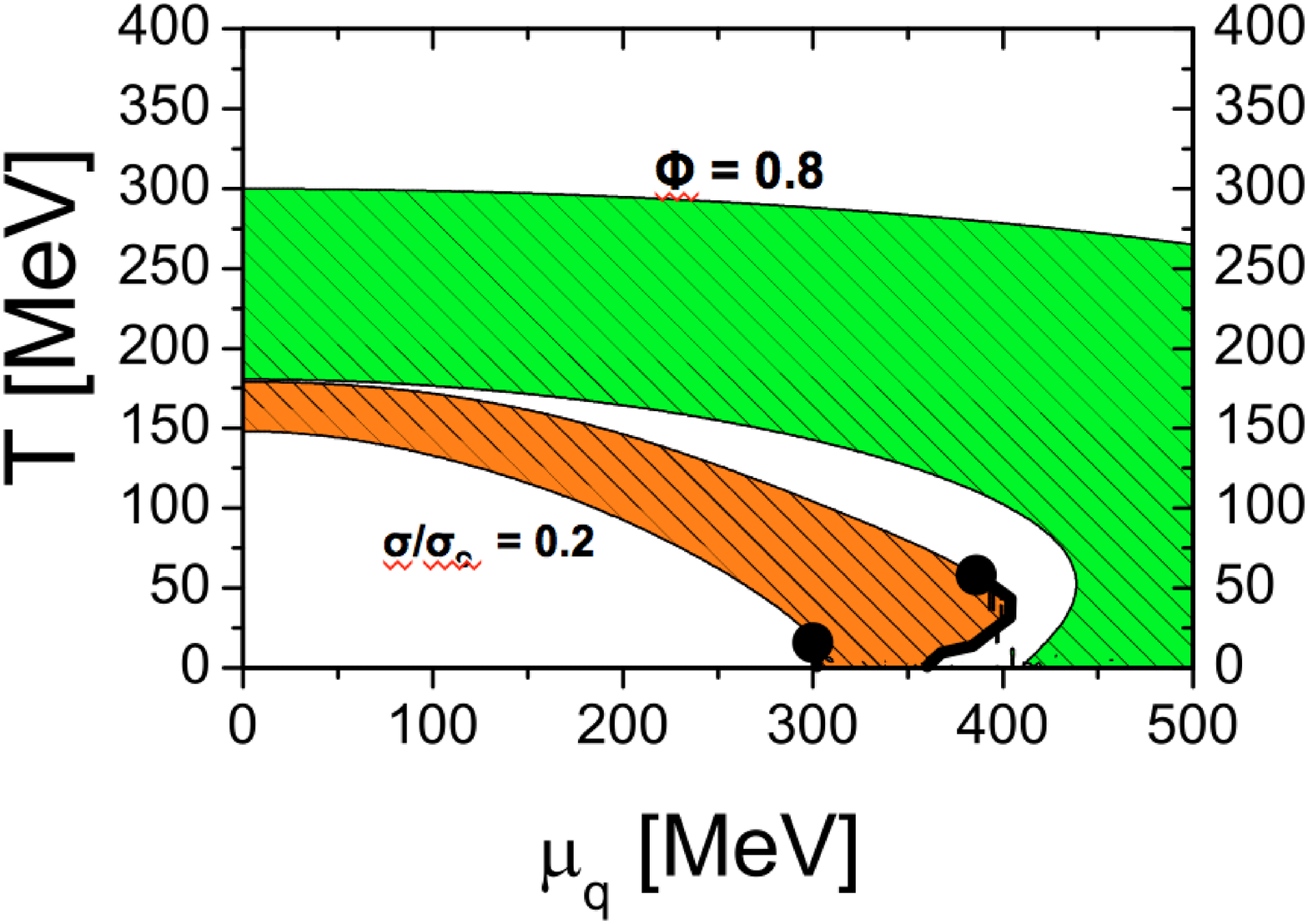}\includegraphics[width=5.8cm]{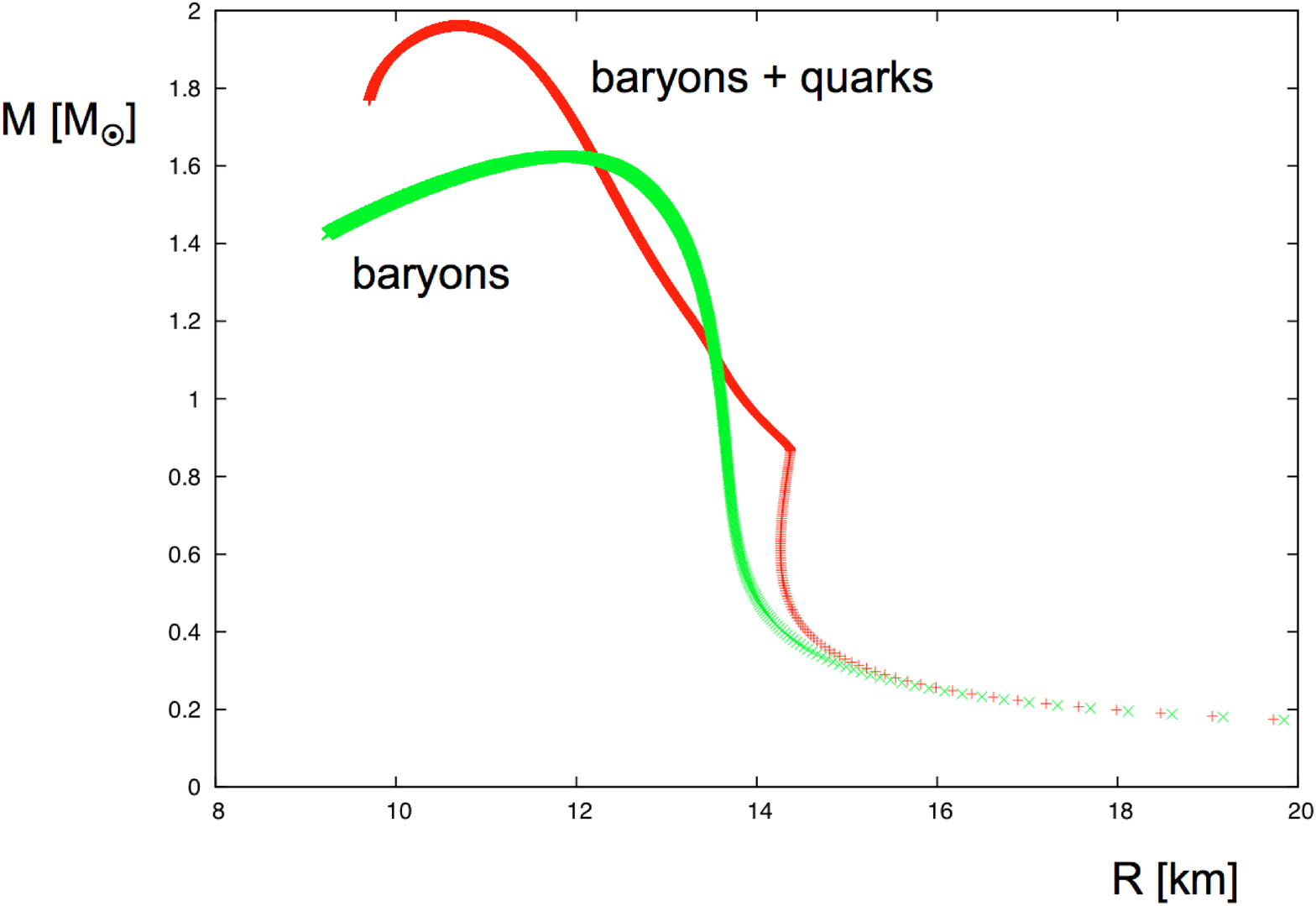}}
\vspace*{5pt}
\caption{Left panel:
Phase diagram of the quark-hadron model as function of temperature and quark chemical potential. 
First-order liquid-gas and chiral transitions are marked by full lines. The lower shaded area represents the range, wherein the scalar field drops from 80 to 20 percent
of its vacuum value. The upper band corresponds to values of the Polyakov loop field $\Phi$ between 0.2 and 0.8.
Right panel: Star masses and radii for the quark-hadron parity model and for the purely hadronic parity model. The latter does not take into account excluded volume corrections.}
\label{tmu}
\end{figure}

The phase diagram of the model is shown in the left panel of Fig. \ref{tmu}. 
There are two first-order phase transition lines both ending in a critical end point, corresponding to the usual
liquid-gas transition and to the chiral phase transition. Separately, at higher temperatures and densities the deconfinement transition occurs as 
symbolized by the upper shaded area.
At vanishing chemical potential the model describes the temperature dependence of
the thermodynamical quantities and order parameters as seen in lattice calculations very well \cite{Steinheimer:2011ea}.
Calculating stars with this approach, one obtains a mass-radius diagram as shown in the right panel of Fig. \ref{tmu} .
The figure contains curves for the purely hadronic and the quark-hadron model. In the case of the hadronic stars no excluded volume
correction was taken into account, as such a term in purely hadronic models inevitably generates acausal speeds of sound larger than the speed of light. 
This is not the case in the QH model as at high densities the system switches to quarks (that do not have an excluded volume) before such conditions can be reached.
In the quark-hadron case two-solar mass stars are possible as can be inferred from the figure. 

\section{Effects of Rotation and the Relevance of Cooling}

\begin{figure}[th]
\centerline{\includegraphics[width=6cm]{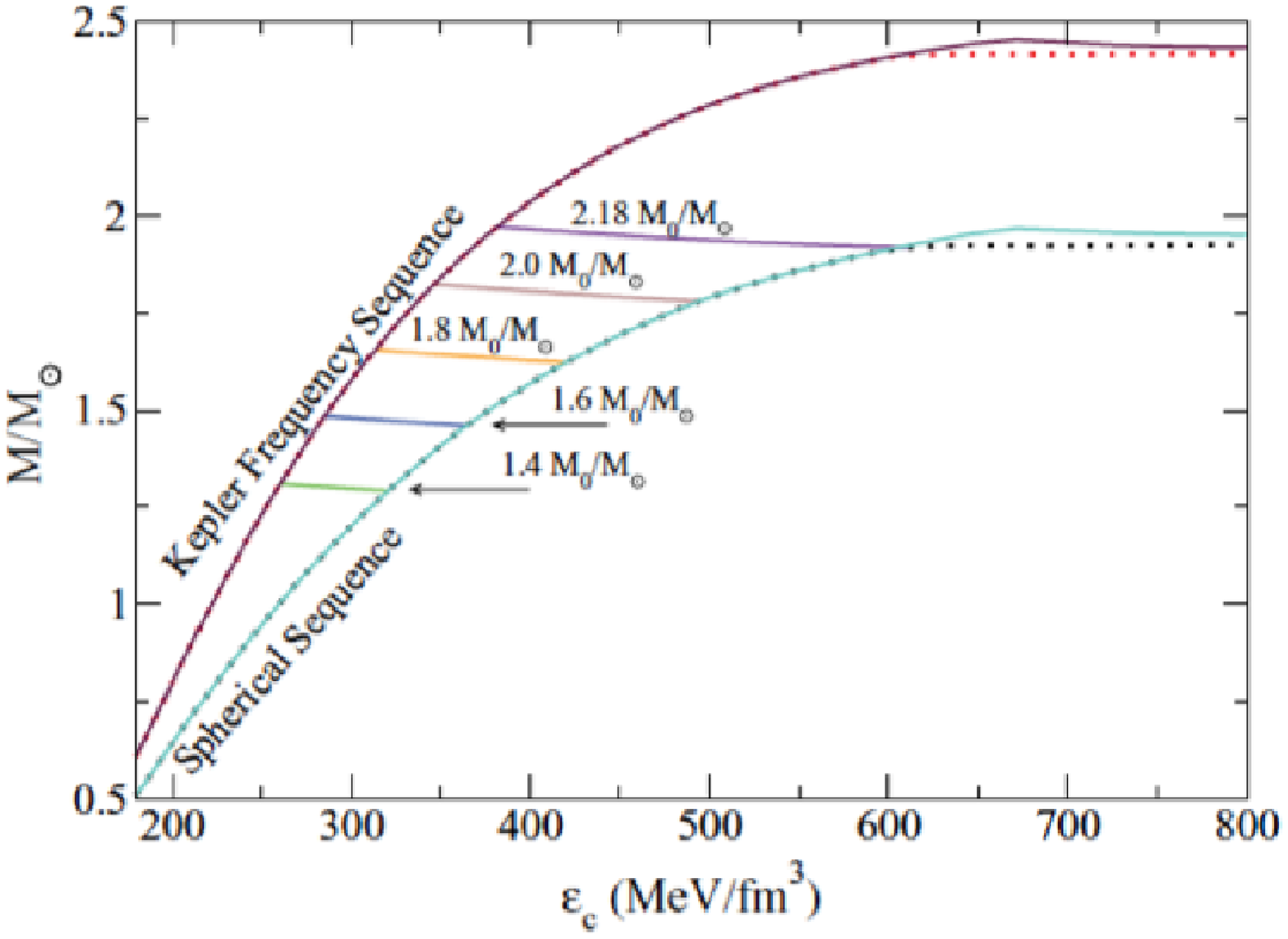}\includegraphics[width=6cm]{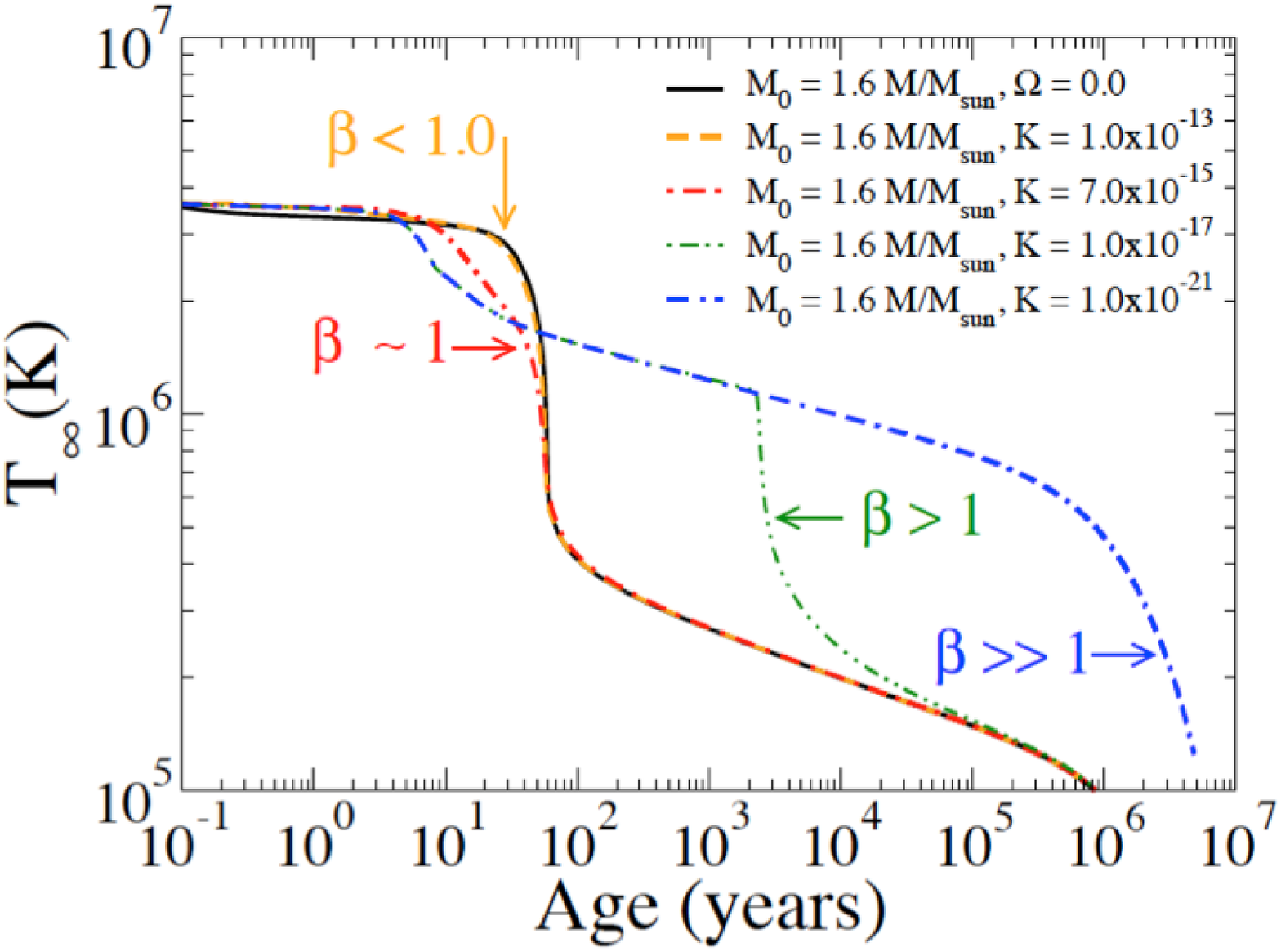}}
\vspace*{5pt}
\caption{Left: Star masses as function of central energy density. Results for the static solutions and for stars rotating at their Kepler frequency
are shown. The nearly horizontal lines indicate the gravitational mass of slowing-down stars at a fixed baryon number.
Right: Cooling curves of neutron stars depending on the ratio $\beta$ of the time scale of the thermal core-crust coupling and spin down}
\label{Afixed}
\end{figure}

Neutron stars can rotate at very high frequencies. Currently the fastest  known rotator is the pulsar  PSR J1748-2446ad with a rotational frequency of 716 Hz \cite{Hessels:2006ze}.
As rotation leads to a decrease of the central density of the star, rotating stars can support larger masses than in the static case as seen in Fig. \ref{Afixed} (left panel).
Following the nearly horizontal lines in the figure corresponds to a spinning down from the maximum (Kepler) frequency to zero rotation at a constant
baryon number. If there is no sizable accretion or mass loss these lines are the evolutionary paths of a slowing-down pulsar.
It is very interesting to note that although the gravitational masses change very little, the central energy density increases by typically 50 percent.
This has a big impact on the structure of the star and can lead to a late appearance of hyperons or quarks in the star while it is spinning down.
Furthermore, combining the effects of rotation and  the cooling of the star can lead to intriguing results. This can be seen in the right panel of Fig. \ref{Afixed}, where the temperature evolution of the neutron
star with time is shown. Depending on the mass of the star a rapid drop of the temperature occurs after about 100 years. 
The reason for this is that the core of the star cools mainly by neutrino emission via direct Urca processes like $n \rightarrow p + e^- + \bar{\nu}_e$,
which are only possible at higher densities of about 2 times saturation density or more (depending on the specific equation of state).
Therefore, cooling occurs mainly in the core of the star.
The cooling wave travels outward and reaches the surface after about 100 years.This is true for the static star. In the case of a rotating star this drop can be delayed
as the star first has to slow down in order to reach inner densities such that the direct Urca process can take place. This is parametrized by the quantity
$\beta$, the ratio of the spin-down
to the core-crust coupling time (in the static case). As the figure shows, for $\beta$ larger than 1 a delayed temperature drop can be observed \cite{Negreiros:2011ak}.
This effect might help to explain the observed cooling evolution of the neutron star in Cas A, that shows a rather steep drop of temperature
300 years after the supernova explosion. This could be understood as the result of the onset of the irect Urca process that was delayed due to rotation as outlined before 
(for details, see \cite{Negreiros:2011ak}).


\section{Conclusions, Outlook}
\label{sec:3}
We studied the properties of compact stars and the phase diagram of strongly interacting matter. 
To that end we extended our hadronic chiral SU(3) model to include quarks.
In addition to study compact stars, in such an approach one can also obtain a very reasonable
description of hot, low-density matter as it is produced in heavy-ion collisions.

The resulting star masses are in agreement with recent observations.
However, as we have discussed, depending on the degrees of freedom taken into account
the inner composition can vary quite substantially without modifying the masses too severely.
Therefore additional observables that are sensitive to the inner structure of the star are important.
Here the study of neutron star cooling is important, as for at least the first thousand years the cooling is neutrino-dominated. The neutrinos
originate mainly in the dense core of the star. Therefore the cooling behavior directly depends on the properties of the constituents in the core
of the star. Especially in conjunction with rotational effects this can lead to new effects as discussed above, that might be tested against observational data.
Here, it is important to perform a full two-dimensional simulation of the cooling of the star. Work along this line is in progress \cite{Negreiros:2012aw}.

\begin{acknowledgement}
We acknowledge the use of the CSC computer facilities at Frankfurt university for our work. R. N. acknowledges financial support from the LOEWE program HIC for FAIR.
T. S. is supported by the Nuclear Astrophysics Virtual Institute (NAVI).
\end{acknowledgement}

\end{document}